\documentstyle[sprocl,epsfig,pstricks]{article}
\input{psfig}

\newcommand{\tr}{\mbox{Tr}}
                           
\let\e=\varepsilon                         
   \let\m=\mu                      
\let\n=\nu   \let\r=\rho \let\s=\sigma                      
                                                      
  \let\PH=\Phi

\let\hc=\dagger

\let\nn=\nonumber                                                               
\def\bea{\begin{eqnarray}} \def\eea{\end{eqnarray}}                             
\def\beann{\begin{eqnarray*}} \def\eeann{\end{eqnarray*}}                       
\def\beq{\begin{equation}} \def\eeq{\end{equation}}                             
\begin{document}

DESY 97-159
\vspace{1cm}
\title{CROSSOVER AND MASS GAP OF THE\\
SU(2)-HIGGS MODEL AT HIGH TEMPERATURE\footnote{Talk given at the Conf.
Strong and Electroweak Matter '97, Eger, Hungary, May 1997}}
\author{W. Buchm\"uller}
\address{Deutsches Elektronen-Synchrotron DESY, Hamburg, Germany}

\maketitle\abstracts{Analytic estimates of screening masses in the symmetric,
high-temperature phase of the SU(2)-Higgs model are reviewed. The size of the
mass gap in the symmetric phase is closely related to the critical Higgs mass
where the first-order electroweak phase transition changes to a smooth crossover.
We also discuss some conjectures concerning the connection between the
screening masses of propagators in a fixed gauge and screening masses of 
gauge-invariant operators.}

The electroweak phase transition \cite{kirzh} is of great cosmological 
significance because baryon number and lepton number violating processes 
are in thermal equilibrium at temperatures above the critical temperature of 
the transition \cite{kuzmin}. In recent years the thermodynamics of the
phase transition has been studied in detail by means of perturbation theory
and numerical lattice simulations. The essential non-perturbative aspects
of the transition can be investigated in the pure SU(2)-Higgs model
neglecting the effects of fermions and the mixing between photon and neutral 
vector boson, which can be included perturbatively. For the dimensionally
reduced theory lattice simulations for the SU(2)xU(1)-Higgs model have been
carried out. We now know that 
the transition is first-order for Higgs boson masses below 70 GeV, and 
that around $\sim 80$ GeV the first-order transition changes to a smooth
crossover\footnote{For recent reviews, see \cite{jan}.}.

The electroweak transition is influenced by non-perturbative effects
whose size is characterized by a `magnetic screening length', the inverse of a
`magnetic mass'. In perturbation theory a magnetic mass appears as a
cutoff which regularizes infrared divergencies \cite{linde}. 
The size of this cutoff is closely related to the confinement
scale of the effective three-dimensional theory which describes the
high-temperature limit of the SU(2) Higgs model \cite{jak}. In the following
we shall review estimates of the magnetic mass by means of gap equations. 
In particular, we shall discuss the connection between the size of the magnetic 
mass and the critical Higgs mass for the onset of a crossover behaviour. We shall
also consider some conjectures concerning the connection between the magnetic
screening mass and screening masses of gauge-invariant operators.  

\section{Gap equations for the magnetic mass}

Consider the SU(2) Higgs model in three dimensions which is defined 
by the action
\beq\label{l3d}
S = \int d^3x \; \tr \left({1\over 2}W_{\mu\nu}W_{\mu\nu} + 
(D_{\mu}\PH)^\hc D_{\mu}\PH + \mu^2 \PH^\hc \PH 
+ 2 \lambda (\PH^\hc \PH)^2 \right) \, , 
\eeq
with 
\beq
\PH = {1\over 2} (\s + i \vec{\pi}\cdot \vec{\tau}) \, ,\quad 
D_{\mu}\PH = (\partial_{\mu} - i g W_{\mu})\PH\, ,\quad  
W_{\mu} = {1\over 2}\vec{\tau}\cdot \vec{W_{\mu}}\ .
\eeq
Here $\vec{W_{\mu}}$ is the vector field, $\s$ is the Higgs field, $\vec{\pi}$
is the Goldstone field and $\vec{\tau}$ is the triplet of Pauli matrices.
The gauge coupling $g$ and the scalar coupling $\lambda$ have mass dimension 
1/2 and 1, respectively. For perturbative calculations gauge fixing and 
ghost terms have to be added. The parameters of the 3d Higgs model are related 
to the parameters of the 4d Higgs model at finite temperature 
by means of dimensional reduction \cite{jak},
\bea
g^2 &=& \bar{g}^2(T) T \, , \quad
\lambda=\bar{\lambda}(T)T + \ldots\, , \\
\mu^2 &=& \left({3\over 16}\bar{g}^2(T) + {1\over 2}\bar{\lambda}(T)\right)
        (T^2-T_b^2)+\ldots \, ,
\eea
where $T$ is the temperature and $\bar{g}$, $\bar{\lambda}$ and $T_b^2$ are
the parameters of the zero temperature 4d theory. Note, that a variation of 
temperature in the 4d theory corresponds to a variation of the parameter $\m^2/g^4$
in the effective 3d theory whereas the ratio $g^2/\lambda$ stays constant.

We are interested in the propagators $G_\s$ and $G_W$ of Higgs field
and vector field, respectively, which at large separation $|x-y|$
fall off exponentially,
\bea\label{expo}
G_\s(x-y) &=& \left\langle \s(x)\s(y)\right\rangle \sim e^{-M |x-y|}\ ,\nn\\
G_W(x-y)_{\m\n} &=& \langle W_{\m}(x)W_{\n}(y)\rangle \sim e^{-m |x-y|}\ .
\eea 
For $\m \gg g^2$ one has $M \simeq \m$, whereas $m$ cannot be computed in 
perturbation theory. An estimate for the vector boson mass can be 
obtained from a coupled set of gap equations for Higgs boson and vector boson 
masses as follows \cite{buphi}. One shifts the Higgs field $\s$ around its vacuum 
expectation value $v$, $\s = v + \s'$, which yields the tree level masses
\beq
m_0^2 = {g^2\over 4}v^2\ ,\ M_0^2 = \mu^2 + 3\lambda v^2.
\eeq
The masses $m_0^2$ and $M_0^2$ are now expressed as
\beq\label{masses}
m_0^2 = m^2 - \delta m^2\ ,\ M_0^2 = M^2 - \delta M^2\ ,
\eeq
\begin{figure}[t]
\begin{center}
\hspace*{0.8cm}
\epsfig{file=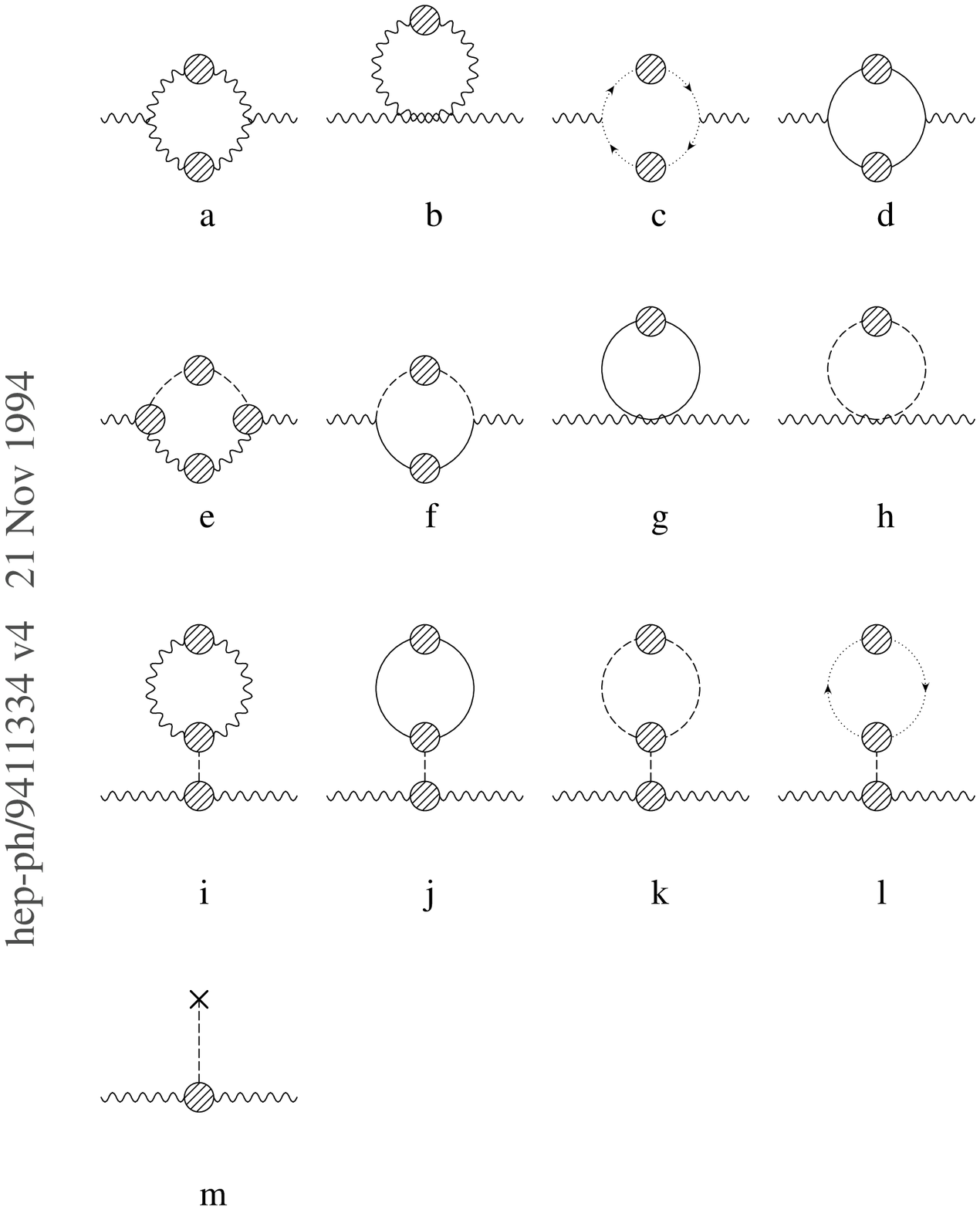,bbllx=68pt,bblly=160pt,%
        bburx=490pt,bbury=740pt,
        width=6cm,clip=}
\end{center}
\caption{\label{fig1} One-loop contributions to the vector boson propagator.
The blobs denote resummed propagators and vertices.}
\end{figure}
where $m$ and $M$ enter the propagators in the loop expansion, and
$\delta m^2$ and $\delta M^2$ are treated perturbatively as counter terms. 
Together with the mass resummation a vertex resummation is performed.
One then obtains a coupled set of gap equations for Higgs boson and 
vector boson masses (cf.~Fig.1),
\bea\label{gaps}
\delta m^2 + \Pi_T(p^2 = -m^2, m, M, \xi) = 0\ ,\nn\\
\delta M^2 + \Sigma(p^2 = -M^2, m, M, \xi) = 0\ ,
\eea
where $\Pi_T(p^2)$ is the transverse part of the vacuum polarization tensor.
The calculation has been carried out in  $R_{\xi}$-gauge. In order to
obtain masses $M$ and $m$ which are independent of the gauge parameter
$\xi$, it is crucial to perform a vertex resummation in addition to
the mass resummation and to evaluate the self-energy terms on the mass shell
\cite{reb}. This yields the screening lengths defined in Eq.~(\ref{expo}).
`Magnetic masses' defined at zero momentum are gauge-dependent \cite{bfhw}.

\begin{figure}[t]
\begin{center}
\hspace*{0.4cm}
\rotateright{
\epsfig{file=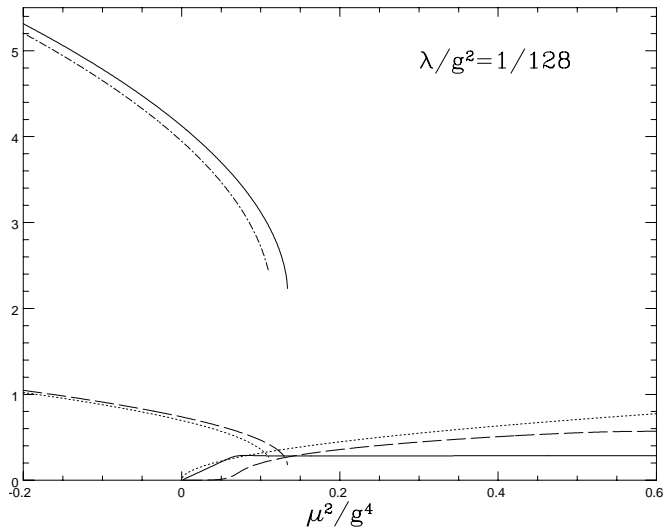,bbllx=47pt,bblly=80pt,%
        bburx=585pt,bbury=747pt,width=7cm,clip=}}
\end{center}
\caption[]{\label{fig2} Vector boson and Higgs boson masses for $\lambda/g^2=1/128$.
Gap equations: m (full line), M (dashed line); perturbation theory: m 
(dash-dotted line), M (dotted line). From Ref.~\cite{buphi}.}
\end{figure}

Together with a third equation for the vacuum expectation value $v$, 
determined by the condition $\langle\s'\rangle = 0$, the gap equations
determine Higgs boson and vector boson masses for each set of values
$\m^2/g^4$ and $\lambda/g^2$. For negative $\m^2$ one finds a unique
solution, corresponding to the familiar Higgs phase, with masses close
to the results of perturbation theory (cf.~Fig.~2). In the case of small
positive $\mu^2/g^4$ and sufficiently small $\lambda/g^2$ there exist
two solutions, corresponding to the Higgs phase and the symmetric phase
with a small, but finite vector boson mass, respectively. This is the 
metastability range characteristic for a first-order phase transition. For large 
positive $\m^2/g^4$ only the solution corresponding to the symmetric
phase remains. Here the Higgs boson mass is $M \simeq \m$, and the
vector boson mass, which is rather independent of $\m$ and $\lambda$,
is given by
\bea\label{msm}
m_{SM} &=& C g^2\ ,\nn\\ 
C &=& {3\over 16\pi}\left({21\over 4} \ln 3 - 1\right) \simeq 0.28 \ .
\eea
In the symmetric phase, where ordinary perturbation theory breaks down,
the vector boson mass is dominated by the graphs a-d in Fig.~1, which 
correspond to the one-loop contributions of the non-linear SU(2)-Higgs 
model \cite{buphi}. Note, that the mass gap is a direct consequence of the 
non-abelian gauge interactions. In the abelian Higgs model no vector boson 
mass is generated in the symmetric phase \cite{buphi2}.

The general strategy and the status of attempts to calculate the mass gap in
the 3d SU(N)-gauge theory by means of gap equations has recently been discussed
by Jackiw and Pi \cite{japi2}. The starting point is always a resummation which
is generated by adding and subtracting an auxiliary action $S_m$ from the
Yang-Mills action $S_G$,
\beq
S_{eff} = {1\over l}\left(S_G(\sqrt{l}W) + S_m(\sqrt{l}W)\right)-
          S_m(\sqrt{l}W)\, .
\eeq
In a perturbative expansion the coefficient of $l^n$ gives the contribution
of all n-loop graphs. $S_m$ is some conveniently chosen, gauge-invariant
mass term. The one-loop gap equation reads
\beq
\Pi_T^{1-loop}(p^2=-m^2) = - m^2\, .
\eeq

In the case of the non-linear $\s$-model, $S_m = S_{\s}$, the calculation
can be considerably simplified. The functional integral for the partition
function is
\beq
Z = \int DW D\pi \Delta\ \exp{-{1\over l}\left(S_G + S_{\s}+S_{GF}
-l S_{\s}\right)}\, 
\eeq
where $S_{GF}$ is some gauge fixing term and $\Delta$ is the corresponding
Faddeev-Popov determinant. The integral over Goldstone fields and ghost
fields can be carried out exactly yielding a massive Yang-Mills
theory without any additional gauge fixing terms \cite{japi2},
\bea
Z &\propto& \int DW \ \mbox{exp}\left[-{1\over l}
\left(S_G(\sqrt{l}W)  
- m^2 \tr\int d^3x \sqrt{l}W_\m \sqrt{l}W_\m\right.\right. \nn \\
&&\hspace{3cm}\left.\left. -l m^2 
\tr\int d^3x \sqrt{l}W_\m \sqrt{l}W_\m \right)\right]\, .
\eea
The exact gap equation now reads
\beq\label{exact}
\Pi_T(p^2=-m^2) = - m^2\, .
\eeq 
Massive Yang-Mills theory corresponds to the non-linear $\s$-model in
unitary gauge. One may therefore expect that the solution of Eq.~(\ref{exact})
agrees with Eq.~(\ref{msm}), whereas the off-shell self energies should only agree
in the limit $\xi \rightarrow \infty$. This is indeed the case \cite{japi2}.

For the 3d SU(N) gauge theory also other gap equation have been considered which
are based on the Chern-Simons eikonal \cite{nair} and on the non-local action 
\cite{jackiw}
\beq\label{nloc}
S_m^{J\cdot P} = m^2 \tr\int d^3x F_{\m}{1\over D^2}F_{\m}\, ,
\eeq
where $F_{\m}= {1\over 2}\e_{\m\n\r}F_{\n\s}$. Amazingly, the one-loop gap
equation of Alexanian and Nair \cite{nair} yields a magnetic mass closely 
related to the one obtained for the non-linear $\s$-model,
\beq
m_{A\cdot N}= {N g^2\over 8\pi} \left({21\over 4}\ln{3} - 1\right) =
              {4\over 3} m_{SM}\, .
\eeq
The choice $S_m^{J\cdot P}$ yields a complex `magnetic mass',
\beq
m_{J\cdot P}= {N g^2\over 8\pi}\left[\left({116\over 16}\ln{3} - {67\over 12}\right)
                \pm i \pi {13\over 16}\right]\, .
\eeq
Note, that Eq.~(\ref{nloc}) can be modified such that the generated mass gap
is real \cite{japi2}.

The evaluation of a magnetic mass by means of gap equations is not a systematic
approach to determine the mass gap in the 3d SU(N) gauge theory since the loop
expansion does not correspond to an expansion in a small parameter. This is
an obvious shortcoming. On the other hand, it may very well be that the 
the one-loop results for $m_{SM}$ and $m_{A\cdot N}$ represent reasonable
approximations of the true mass gap. In this respect an extension of the
gap equations to two-loop order would be very valuable \cite{eber}. If the whole 
approach makes sence the two-loop correction should be of order 
$m_{SM}-m_{A\cdot N}$. It is also encouraging that the magnetic masses $m_{SM}$ and
$m_{A\cdot N}$ are consistent with the propagator mass obtained in a numerical 
lattice simulation in Landau gauge \cite{neu},
\beq
m_{SM}^{(L)} = 0.35 (1) g^2\, .
\eeq
The lattice simulation also yields a dependence of the vector boson mass on the
parameter $\m^2/g^4$ which is very similar to the solution of the gap equation
shown in Fig.~3.

\section{Mass gap and crossover}

A direct consequence of the mass gap in the symmetric phase is the change of
the first-order phase transition at small Higgs masses to a smooth crossover
at some critical Higgs mass $\bar{m}_H^c$. Fig.~3 shows the solution of the
gap equations for a physical Higgs mass equal to the W-boson mass of about
80 GeV. The solution is very different from the one shown in Fig.~2 which
corresponds to a Higgs mass of about 20 GeV. In Fig.~3 the variation of vector
and scalar screening masses is continuous as $\m^2/g^4$ is increased from the 
region of the Higgs phase to the domain of the symmetric phase. Based on the
gap equations it was predicted that the critical Higgs mass $\bar{m}_H^c$
should be below 100 GeV \cite{buphi}. Recently, convincing evidence for a crossover 
has indeed been obtained in numerical lattice simulations, and the critical Higgs
mass has been determined to be about 80 GeV  \cite{lai,ran}.

\begin{figure}
\begin{center}
\hspace*{0.4cm}
\rotateright{
\epsfig{file=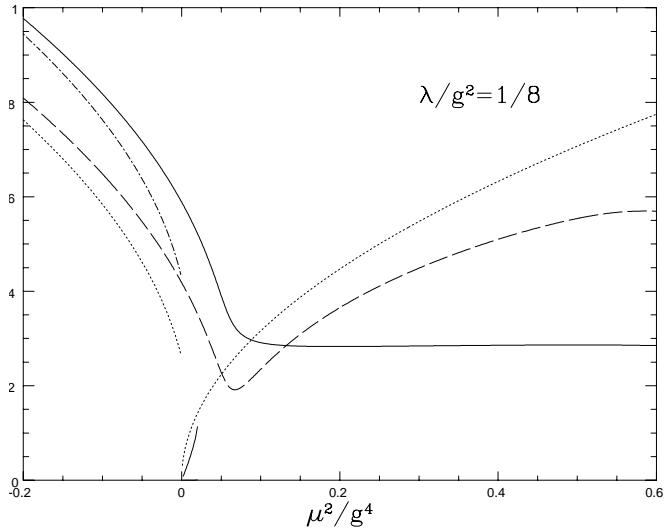,bbllx=47pt,bblly=80pt,%
        bburx=585pt,bbury=747pt,width=7cm,clip=}}
\end{center}
\caption[]{\label{fig3} Vector boson and Higgs boson masses for $\lambda/g^2=1/8$
which corresponds to a physical Higgs mass equal to the W-boson mass. Gap equations:
m (full line), M (dashed line); perturbation theory: m (dash-dotted line), M
(dotted line). From Ref.~\cite{buphi}.}
\end{figure}

In the SU(2)-Higgs model a crossover behaviour was first observed for the 
finite-temperature 4d theory in numerical lattice simulations for large Higgs 
masses by Evertz, Jers\'ak and Kanaya, who also discussed in detail the phase 
diagram \cite{dam}. Also in the average action approach to the electroweak
phase transition a change from a first-order transition to a crossover is
expected \cite{bewe1,bewe2,tet}. In this framework, information about 
non-perturbative properties of the theory from elsewhere is at present needed 
to obtain an estimate for the critical Higgs mass which could be as large as
200 GeV or larger \cite{bewe1}. The general argument, that three-dimensional 
confinement forbids massless particles and therefore a second-order transition 
\cite{reu}, is at variance with the observed change from a first-order transition 
to a crossover and the general expectation that a first-order critical line 
ends in a second-order transition. It is important to clarify the nature of the
endpoint for the electroweak transition.

The essence of the connection between  mass gap and critical Higgs boson mass
in the gap equation approach can be easily understood as follows \cite{buphi3}. 
Consider the one-loop effective potential in unitary gauge ($\vec{\pi} = 0$), 
\beq\label{v1l}
V_{1l} = {1\over 2} \m^2 \s^2 + {1\over 4}\lambda \s^4
         - {1\over 16\pi} g^3 \s^3\, ,
\eeq
where we have neglected the scalar contributions for simplicity.
At the beginning of the metastability range, $\mu^2=0$, the Higgs vacuum
expectation value is 
$\s_0 = 3g^3/(16\pi \lambda)$, which corresponds to
the vector boson mass
\beq
m_W (\m^2 = 0) = {3\over 32 \pi} {g^4\over \lambda}\, .
\eeq
It is reasonable to expect that the first-order phase transition dissappears
at a critical scalar coupling where the vector boson mass in the Higgs phase 
reaches the magnetic mass of the symmetric phase. The condition 
$m_W (\m^2 = 0) = m_{SM}$
determines a critical coupling $\lambda_c$. The corresponding 
zero-temperature critical Higgs boson mass is given by 
\beq\label{crith}
\bar{m}^c_{H} = \left({3\over 4\pi C}\right)^{1/2}\bar{m}_W
              \simeq 74\ \mbox{GeV}\, ,
\eeq
where $\bar{m}_W$ is the zero-temperature vector boson mass. Eq.~(\ref{crith})
clearly shows that the crossover point is determined by the constant $C$,
i.e., the size of the magnetic mass. 
The obtained value of 
the critical Higgs mass agrees rather well with the result of recent 
numerical simulations \cite{lai,ran}. In contrast, for vanishing magnetic mass the
first-order transition never changes to a crossover, while taking $C>1.0$ 
corresponding to the measured bound state mass $m_V$ (cf.~Eq.~(\ref{vec}), 
table~1) grossly underestimates $\bar{m}_H^c$.

\begin{figure}[t]
\vspace{-1cm}
\begin{center}
\leavevmode
\epsfysize=300pt
\epsffile{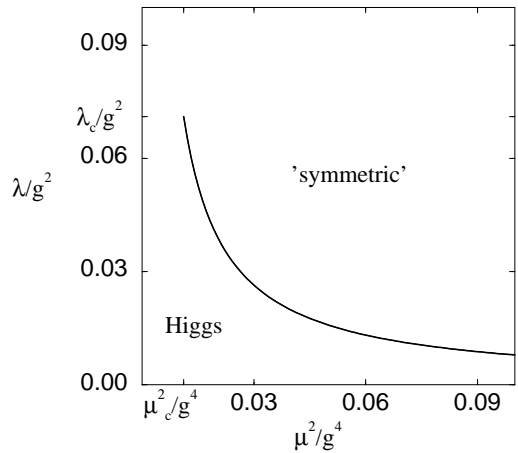}
\vspace{-4cm}
\end{center}
\caption[]{\label{phased}\it
Critical line of first-order phase transitions as given by Eq.~(\ref{crit}).
From Ref.~\cite{buphi3}.}
\vspace*{0.5cm}
\end{figure} 

From the effective potential (\ref{v1l}) one can easily determine the
critical line of the first-order phase transition. For 
$\hat{\lambda}(\m^2) < \lambda_c$ the conditions
\beq
0 = V_{1l}(\s_0)
  = \left.{\partial\over \partial \s} V_{1l}\right|_{\s=\s_0}
\eeq
yield for the critical line $\hat{\lambda}(\m^2)$,
\beq \label{crit}
{\hat{\lambda}(\m^2)\over g^2} = {1\over 128\pi^2} {g^4\over \mu^2}\ .
\eeq
The corresponding phase diagram is shown in Fig.~\ref{phased}. 
The critical value of 
the mass parameter at the crossover point is given by
\beq
{\m_c^2\over g^4} \simeq {C\over 8\pi}\, .
\eeq 
Using the matching relations to the finite-temperature Higgs model one
can evaluate the critical temperature as function of the zero-temperature
Higgs boson mass.\\

\section{Gauge-invariant correlation functions}

It is known that the SU(2) Higgs model has only a single phase, and
that the Higgs and the confinement regime are analytically connected. 
All physical properties of the model can be obtained by studying
correlation functions of gauge-invariant operators. This is of particular
importance for numerical lattice simulations where in general the gauge is 
not fixed.

In the literature the following operators for scalar states with 
$J^{PC}=0^{++}$ have been studied,
\bea
R(x) &=& \tr\left(\PH^{\hc}(x)\PH(x)\right)\, ,\\
L(x) &=& \tr\left((D_{\m}\PH)^{\hc}D_{\m}\PH(x)\right)\, ,\\
P(x) &=& {1\over 2} \tr\left(W_{\m\n}W_{\m\n}\right)
      = - {1\over 8g^2}\tr\left([D_{\m},D_{\n}][D_{\m},D_{\n}]\right)\, .
\eea
The standard operator for vector states with $J^{PC}=1^{--}$ is
\beq\label{vec}
V^a_{\mu}(x) = {1\over 2} 
\tr\left(\PH^{\hc}(x)\stackrel{\leftrightarrow}{D_{\m}}\PH(x)\tau^a\right)\, .
\eeq
In the numerical simulations \cite{mon,kajla,tep,neu,gue} screening masses 
have been determined from the 2-point functions of the operators $R$ and 
$V^a_{\m}$,
\bea
G_R(x-y) &=& \langle R(x) R(y) \rangle \sim  e^{-m_R |x-y|}\, ,\\
G_V(x-y)_{\m\n} &=& \langle V_{\m}(x) V_{\n}(y) \rangle \sim  e^{-m_V |x-y|}\, .
\eea
In \cite{tep} screening masses have also been measured for the
operators $L(x)$ and $P(x)$,
\bea
G_L(x-y) &=& \langle L(x) L(y)\rangle \sim e^{-m_L |x-y|}\, ,\\
G_P(x-y) &=& \langle P(x) P(y)\rangle \sim e^{-m_P |x-y|}\,.
\eea
The screening masses $m_R$, $m_V$, $m_L$ and $m_P$ have been determined
for positive and negative values of $\m^2$, i.e., in the symmetric phase
and in the Higgs phase. In the latter case, as expected, the results agree
with perturbation theory and gap equations. The value for $m_P$ is 
consistent with an intermedate state of two massive vector bosons $V$
contributing to $G_P$. This is the leading contribution if one expands
$P(x)$ in powers of $g^2$.

In the symmetric phase, however, the numerical results for $m_R$ and $m_V$
do not agree with the predictions of the gap equations. Since also in the
symmetric phase the vacuum expectation value of the Higgs field is
different from zero, it was suggested in \cite{buphi} that the magnetic
mass $m_{SM}$ should determine the asymptotic behaviour of $G_V$ 
(cf.~(\ref{fms})). The numerical simulations show no sign of this.

What is the connection between the gauge-dependent 2-point function
$G_W$ and the gauge-invariant 2-point function $G_V$? This question has
been addressed by Fr\"ohlich, Morchio and Strocci in their  
detailed study of the Higgs phenomenon in terms of gauge-invariant
operators \cite{froe}. As they have pointed out,  gauge-invariant
correlation functions are 
approximately proportional to gauge-dependent correlation
functions as calculated in standard perturbation theory, if for the
chosen gauge and renormalization scheme the fluctuations of the Higgs
field are small compared to the vacuum expectation value. For instance,
for the scalar correlation functions one has ($\s=v+\s', 
\langle \s'\rangle =0$),
\beq\label{fms}
\langle R(x) R(y)\rangle \sim v^2 \left(\langle \s'(x) \s'(y)\rangle
 + \mbox{\cal O}\left({\s' \over \langle \s \rangle}, {\vec{\pi}\over 
        \langle \s\rangle}\right)\right)\, .
\eeq
As a measure for the relative size of the fluctuation terms one may consider 
the ratio \cite{buphi3}
\beq\label{ratio}
\zeta = {\langle \PH^{\hc}\PH \rangle \over \langle \s \rangle^2}\, .
\eeq
At one-loop order one obtains in $R_{\xi}$-gauge,
\bea\label{fluc}
\langle \PH^{\hc}\PH\rangle &=& \langle \s^2+\vec{\pi}^2\rangle
 = v^2 + \langle \s'^2 + \vec{\pi}^2\rangle \\
&=& v^2 - {1\over 4\pi}\left(M + 3 \sqrt{\xi} m \right)\, .
\eea
Here linear divergencies have been subtracted by means of dimensional
regularization. Deep in the Higgs phase, where $\m^2 < 0$, 
$v_0^2 = - \m^2/\lambda$,
$M^2_0 = 2\lambda v_0^2$ and $m_0^2 = g^2 v_0^2/4$, one finds
\beq
\zeta_H = 1 - {3\over 8\pi}\left(\sqrt{\xi} + 
   {2\over 3} {\sqrt{2\lambda}\over g}\right) {g\over v} + \ldots\, .
\eeq
In the relevant range of parameters one has $g/v < 1$. Hence, the deviation 
of $\zeta_H$ from 1 is small and ordinary perturbation theory is reliable. 

On the contrary, in the symmetric phase the situation is very different.
Here the gap equations yield for the vacuum expectation value $g/v
\simeq 10$. Inserting in the definition of the ratio (\ref{ratio}) solutions 
of the gap equations for $M$ and $m$ one finds in the symmetric phase
that $\zeta_{SM}$ deviates from 1 by more than $100\%$.
Hence, we cannot expect that the gauge-dependent 2-point functions give a 
good approximation to the gauge-invariant 2-point functions. 

\section{Gauge-invariant screening masses}
                  
\begin{figure}[t]
\vspace{-2cm}
\begin{center}
\leavevmode
\epsfysize=140pt
\hspace*{-2.5cm}
\epsffile{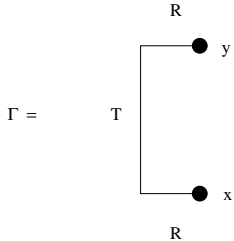}
\vspace{-0.5cm}
\end{center}
\caption[]{\label{gamma}\it The path $\Gamma$ for $G_{\PH}(T,R)$.}
\vspace*{0.5cm}
\end{figure}

What is the physical meaning of the propagator masses obtained from gap 
equations as well as numerical simulations in the symmetric phase? 
Several years ago the notion of a `screening energy' has been introduced
in connection with an analysis
of the SU(2) Higgs model at zero temperature \cite{kas}.
The authors considered the gauge-invariant correlation function
\beq \label{gphi}
G_{\PH}(T,R) = \left\langle\tr\left(\PH^{\hc}(y)U(\Gamma)\PH(x)
                     \right)\right\rangle\, ,
\eeq
where
\beq
U(\Gamma) = P \exp{\left(ig\int_{\Gamma}ds\cdot W\right)}
          \equiv U^{\hc}(R,y)U(T)U(R,x)\, ,
\eeq
and the path $\Gamma$
\beq
\Gamma \equiv \Gamma(y,R)\circ\Gamma(T)\circ\Gamma(R,x)
\eeq
is shown in Fig.~\ref{gamma}.
For large $T$, with $R$ fixed, an exponential fall-off was found,
\beq\label{mcon}
G_{\PH}(T,R) \sim e^{-m_{\PH}T}\, ,
\eeq
with $m_{\PH}$ being independent of $R$. In temporal gauge the 2-point
function takes the form,
\beq\label{ener}
G_{\PH}(T,R) = \left\langle \tr\left(\PH^{\hc}(x)U^{\hc}(R,x)e^{-HT}U(R,x)
               \PH(x)\right)\right\rangle\, ,
\eeq
where $H$ is the hamiltonian. Comparison of Eqs.~(\ref{mcon})
and (\ref{ener}) suggests that $m_{\PH}$ is the energy of a dynamical charge
bound by an external charge \cite{kas}. If the energy of the infinitely
heavy external charge is properly subtracted, $m_{\PH}$ corresponds to the
`constituent', or screening mass of the bound scalar $\PH$. In the case $R=0$ 
the 2-point function $G_{\PH}$ reduces to the gauge-invariant propagator
\beq
\hat{G}_{\PH}(x-y)=\left\langle \tr\left(\PH^{\hc}(y)U_{yx}\PH(x)\right)
                   \right\rangle
                   \sim e^{-m_{\PH}|x-y|}\, ,
\eeq
where $U_{yx}$ is the non-abelian phase factor along the straight line
from $x$ to $y$. 

The definition of a screening mass for the vector boson is completely
analogous. The obvious definiton is
\beq
\hat{G}_W(x-y)_{\m\n\r\s} = \left\langle W^T_{\m\n}(y)
U^{\cal A}_{yx} W_{\r\s}(x)\right\rangle
\sim e^{-m_W |x-y|}\, .
\eeq
where the superscript ${\cal A}$ denotes SU(2) matrices in the adjoint
representation.

The contribution from the phase factor to the masses $m_{\PH}$ and
$m_W$, which depend on the mass parameter $\m^2$, are linearly divergent. 
Renormalized screening masses can be 
defined by matching $m_{\PH}$ and $m_W$ to the masses $m_R$ and $m_V$
of the gauge-invariant correlation function at some value $\m^2_0$ in the 
Higgs phase. The corresponding screening masses $m_{\PH}(\m^2;\m^2_0)$ and
$\m^2_W(\m^2;\m^2_0)$ satisfy the boundary conditions
\beq
m_{\PH}(\m^2_0;\m^2_0) = m_R(\m^2_0)\quad ,\quad
m_W(\m^2_0;\m^2_0) = m_V(\m^2_0)\; .
\eeq
These screening masses, as functions of $\m^2$, should behave similarly as
the solutions $M(\m^2)$ and $m(\m^2)$ of the gap equations, respectively.
\begin{figure}[t]
\begin{center}
\leavevmode
\epsfysize=150pt
\epsfbox[20 400 620 730]{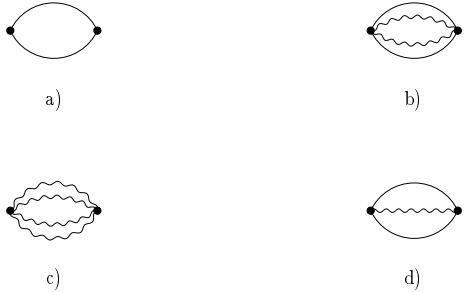}
\vspace{-1.5cm}
\end{center}
\caption[]{\label{2pt}\it
  Two-point functions to leading order for a constituent model.}
\vspace*{0.5cm}
\end{figure}

What is the role of the screening masses in the correlation functions
of gauge-invariant operators? As discussed above, the fluctuations dominate
in the symmetric phase. Hence, one may expect that multi-particle states
of `constituent' scalar and vector bosons dominate the exponential fall-off
of the 2-point functions \cite{buphi3}. This is similar in spirit to the bound 
state model of Dosch et al. \cite{dosch}. According to Fig.~\ref{2pt},
for $G_R$ this should be a $(\PH^{\hc}\PH)$ state (a),
for $G_L$ a $(\PH^{\hc}W W \PH)$ state (b), for $G_P$ a $(W W W W)$ state (c) 
and for $G_V$ a $(\PH^{\hc}W\PH)$ state (d). 
Here we have identified a covariant
derivative $D_{\m}$ with a constituent vector boson $W$, since for bound
states in the symmetric phase an expansion in powers of $g^2$ is not
justified. Neglecting binding effects, this yields the mass formulae
\bea
m_R \simeq 2 m_{\PH}\,\quad m_L \simeq 2 m_{\PH} + 2m_{W}\, 
\quad m_P \simeq 4 m_W\, \quad m_V \simeq 2 m_{\PH} + m_W\, .
\eea 
These relations can be compared with results from lattice simulations.
A screening mass $m_W$ for the vector boson was determined in \cite{neu},
$m_W = 0.35(1) g^2$. No scalar screening mass has been measured so far,
hence we choose $m_{\PH} = m_R/2$. This yields three predictions for
$m_L$, $m_P$ and $m_V$ which are compared with the results of Ref.~\cite{tep}
in table~1. The qualitative agreement supports the constituent picture.
Note, that in the bound state model \cite{dosch} no prediction has so far been
made for the W-ball mass $m_P$.

\renewcommand{\arraystretch}{1.5}
\begin{table}[h]
\begin{center}
\begin{tabular}{|c|cccc|c|}
\hline
    &&$J^{PC}=0^{++}$ & & &$J^{PC}=1^{--}$    \\
    &$ R $ &$ L $ & &$ P $ &$ V $  \\ \hline\hline
lattice simulations \cite{tep}
& 0.839(15) & 1.47(4) & & 1.60(4) & 1.27(6) \\ \hline
constituent model & -  & 1.54$\;$ & & 1.40$\;$ & 1.18$\;$  \\ \hline
\end{tabular}
\end{center}
\caption[]{\it Comparison of screening masses from lattice simulations
and a constituent model. $m_R$ is used to fix the constituent scalar mass. 
From Ref.~\cite{buphi3}.}
\end{table}

The proposed picture can be tested by measuring
the gauge-invariant propagators $\hat{G}_{\PH}$ and $\hat{G}_W$ as
functions of $\m^2$. The masses $m_{\PH}(\m^2;\m^2_0)$ and 
$m_W(\m^2;\m^2_0)$ should 
behave like the solutions $M(\m^2)$ and $m(\m^2)$ of the gap equations.
In particular, at a first-order transition from the Higgs phase to the
symmetric phase, both screening masses should jump to smaller values.
With increasing $\lambda/g^2$ the jump should decrease and eventually
vanish at the critical coupling where the crossover behaviour sets in.

We hope that further numerical and analytical investigations will unequivocally
clarify the relation between the various screening masses and provide a clear
physical picture of the symmetric phase. This will be crucial in order to
understand better real-time correlation functions at high temperature, in
particular the sphaleron rate \cite{arnold}.

\section*{Acknowledgments}
It is a pleasure to thank Owe Philipsen for a continuing collaboration on a very
often confusing topic and to thank the organizers for aranging an enjoyable and 
stimulating conference.

\end{document}